\documentclass[letterpaper, 10 pt, journal, twoside]{ieeetran}

\pdfoutput=1
\usepackage[caption=false]{subfig}
\usepackage{graphicx}
\usepackage{amsmath,amssymb,bm,mathrsfs,xfrac,mathtools,mathrsfs,amsthm}
\newcommand*{\tran}{^{\mkern-1.5mu\mathsf{T}}}
\usepackage{color}
\usepackage[hyphens]{url}
\usepackage[hidelinks]{hyperref}
\usepackage{soul}
\usepackage{cite}
\usepackage[export]{adjustbox}
\usepackage{float}
\usepackage{blindtext}
\usepackage{multirow}
\usepackage{pgf,tikz}
\newcommand\norm[1]{\left\lVert#1\right\rVert} 
\renewcommand{\baselinestretch}{.99} 
\usetikzlibrary{arrows,calc}
\IEEEoverridecommandlockouts

\usepackage{booktabs}
\usepackage{colortbl}
\definecolor{mygray}{gray}{0.9} \sethlcolor{mygray}

\usepackage{color, colortbl}
\definecolor{NREL1}{RGB}{000,166,222}

\newcommand{\tikzcircle}[2][red,fill=red]{\tikz[baseline=-0.5ex]\draw[#1,radius=#2] (0,0) circle;}
\newcommand{\sqdiamond}[1][fill=black]{\tikz [x=1.2ex,y=1.85ex,line width=.1ex,line join=round, yshift=-0.285ex] \draw  [#1]  (0,.5) -- (.5,1) -- (1,.5) -- (.5,0) -- (0,.5) -- cycle;}%
\newcommand{\MyDiamond}[1][fill=black]{\mathop{\raisebox{-0.275ex}{$\sqdiamond[#1]$}}}

\usepackage{dashrule}

\title{On analytical construction of observable functions in extended dynamic mode decomposition for nonlinear estimation and prediction}
\author{Marcos~Netto,~\IEEEmembership{Member,~IEEE,}~Yoshihiko~Susuki,~\IEEEmembership{Member,~IEEE,}~Venkat~Krishnan,~\IEEEmembership{Senior~Member,~IEEE,} and~Yingchen~Zhang,~\IEEEmembership{Senior~Member,~IEEE}
\thanks{M. Netto, V. Krishnan, and Y. Zhang are with the Power Systems Engineering Center, NREL, Golden, CO 80401, USA. Y. Susuki is with the Department of Electrical and Information Systems, Osaka Prefecture University, Osaka 599-8531, and JST, PRESTO, 4-1-8 Honcho, Kawaguchi, Saitama 332-0012, Japan. M. Netto is supported by the Director's Postdoctoral Fellowship under the Laboratory Directed Research and Development program at NREL. Y. Susuki is supported in part by JST, PRESTO Grant No. JP-MJPR1926. V. Krishnan is supported by the U.S. Department of Energy, Office of Electricity, Grant No. DE-OE0000876. Corresponding author: \href{mailto:marcos.netto@nrel.gov}{marcos.netto@nrel.gov}.}}

\begin{document}
\maketitle
\thispagestyle{empty} 

\begin{abstract}
We propose an analytical construction of observable functions in the extended dynamic mode decomposition (EDMD) algorithm. EDMD is a numerical method for approximating the spectral properties of the Koopman operator. The choice of observable functions is fundamental for the application of EDMD to nonlinear problems arising in systems and control. Existing methods either start from a set of dictionary functions and look for the subset that best fits the underlying nonlinear dynamics or they rely on machine learning algorithms to ``learn'' observable functions. Conversely, in this paper, we start from the dynamical system model and lift it through the Lie derivatives, rendering it into a polynomial form. This proposed transformation into a polynomial form is exact, and it provides an adequate set of observable functions. The strength of the proposed approach is its applicability to a broader class of nonlinear dynamical systems, particularly those with nonpolynomial functions and compositions thereof. Moreover, it retains the physical interpretability of the underlying dynamical system and can be readily integrated into existing numerical libraries. The proposed approach is illustrated with an application to electric power systems. The modeled system consists of a single generator connected to an infinite bus, where nonlinear terms include sine and cosine functions. The results demonstrate the effectiveness of the proposed procedure in off-attractor nonlinear dynamics for estimation and prediction; the observable functions obtained from the proposed construction outperform methods that use dictionary functions comprising monomials or radial basis functions.
\end{abstract}


\vspace{-.1cm}
\section{Introduction}
\IEEEPARstart{K}{oopman} operator theory (KOT) and associated numerical methods \cite{Mauroy2020} are promising for system identification \cite{Mauroy2020b}, state estimation \cite{Netto2018b}, stability assessment \cite{Mauroy2016}, and control \cite{Korda2020} of nonlinear dynamical systems. The increasing interest in the applications of KOT to systems and control is primarily because of two of its characteristics: i) it does not rely on any model---from beginning to end, numerical methods based on KOT are truly data driven, yet they are supported by a mathematical foundation anchored on the spectral theory of dynamical systems \cite{Mauroy2020}; and ii) linear and \emph{nonlinear} modes are captured, although the numerical methods rely exclusively on linear algebra. In simple words, these outstanding characteristics can be explained by the fact that the Koopman operator is a linear, infinite-dimensional operator that acts on functions. In principle, any measured quantity of a dynamical system can be expressed as a function of its state variables, $\bm{x}$, hereafter referred to as an observable function, $g(\bm{x})$. See, e.g., \cite{Mauroy2020} for a formal exposition of this topic.

A great deal of progress has been made in devising numerical methods that provide a finite-dimensional approximation to the infinite-dimensional Koopman operator. Extended dynamic mode decomposition (EDMD) \cite{Williams2015} is an example of a powerful numerical method tailored to this purpose; see \cite{Korda2018a} for a study on the convergence of EDMD to the Koopman operator. EDMD is, however, sensitive to the set of observable functions provided as input \cite{Otto2019}; finding the right set of observable functions, e.g., a set of observable functions that yields a Koopman invariant subspace \cite{Brunton2016x}, is nontrivial and an unsolved problem. Although there is an increase in the number of applications based on EDMD, very few researchers have tackled the fundamental challenge of choosing the right set of observable functions. One strategy is to start from a large set of dictionary functions and apply a sparse regression penalty on the number of functions selected to approximate the dynamics of the underlying system \cite{Brunton2016b}. Another strategy is to use neural networks to ``learn'' the observable functions \cite{Li2017, Lusch2018, Otto2019, Yeung2019}. These strategies have found success in a broad range of complex problems where data are abundant and state-space models are scarce or nonexistent. A question that always plagues these strategies is how well the discovered mapping describes the underlying system dynamics \emph{beyond} sampled trajectories. This is a challenging question to answer with limited to no access to a model. On the other hand, for real engineering systems, state-space models are often available in systems and control. Although in certain cases the uncertainty in the parameters is large, the structure of the model is known. In this context, for a given set of observable functions, one can optimize the approximated spectral properties of the Koopman operator, in particular the Koopman eigenfunctions \cite{Korda2018b}. But the selection of observable functions is particularly challenging if the state-space model contains nonlinear terms given by nonpolynomial functions and the underlying dynamical system has multiple fixed points \cite{Brunton2016x}. Note that the Carleman linearization \cite{Kowalski1991} is limited to polynomial vector fields.

The challenge in selecting observable functions is very well explained in \cite{Mamakoukas2019}, wherein time derivatives of the underlying nonlinear dynamical system are used to approximate Koopman operators in the vicinity of fixed points. The rationale of using time derivatives essentially is that of a Taylor series expansion, except that one replaces the nonlinear terms obtained from the time derivatives by observable functions, and then proceeds on the expansion. Note that the lifting of dynamical systems containing terms such as $\sin{x}$ and $\cos{x}$ by using time derivatives has no closure; hence, the set of observable functions obtained by using the method proposed in \cite{Mamakoukas2019} needs to be truncated at a certain point. The problem of selecting a set of observable functions to obtain a linear embedding for engineering systems with known model structure that is representative of the entire domain of attraction has not been addressed before. Moreover, a method that provides a \emph{closed} set of observables functions is nonexistent; these are the main contributions of this letter. We start by noticing that a broad class of dynamical systems \cite{Sigg2014, Grasser2003, Netto2018b, Netto2018c} comprise elementary nonlinear functions, such as $\sin{x}$, $\cos{x}$, $e^{x}$, $\frac{x}{b+x}$, and compositions of these elementary functions; and that dynamical systems that fall into this class can be put into a polynomial form by lifting the original system to a space of higher dimension. The lifting procedure, originally proposed in \cite{Gu2011}, relies on Lie derivatives. We show that this embedding, referred to as \emph{polynomialization}, makes EDMD more effective for the aforementioned class of dynamical systems. This letter is intended to set the direction for others working on the problem of selecting observable functions. Additionally, we envision the numerical illustration in Section \ref{sec.IV} to serve as a benchmark problem and solution that researchers could use to compare their choice of observable functions against.

\vspace{-.1cm}
\section{Preliminaries}

\subsection{Koopman Operator Theory \cite{Mauroy2020}}
Let an autonomous dynamical system evolving on a finite, $n$-dimensional manifold $\mathbb{X}$ be:

\vspace{-.3cm}\begin{equation}
\bm{\dot{x}}(t) = \bm{f}(\bm{x}(t)), \label{eq.1z}
\end{equation}

\vspace{-.1cm}\noindent
for continuous time $t\in\mathbb{R}$, where $\bm{x}\in\mathbb{X}$ is the state, and $\bm{f}:\mathbb{X} \to T\mathbb{X}\,(\text{tangent bundle of}\,\mathbb{X})$ is a nonlinear vector-valued function. In what follows, we introduce the Koopman operator for continuous time systems. Let $g(\bm{x})$ be a scalar-valued function defined in $\mathbb{X}$, such that $g:\mathbb{X}\to\mathbb{C}$. The function $g$ is referred to as an \emph{observable function}. The space of observable functions is $\mathcal{F}\subseteq{C}^{0}$, where ${C}^{0}$ denotes all continuous functions. Note that the choice of $\mathcal{F}$ is discussed in \cite{Mauroy2020}. The Koopman operator, denoted by $\mathcal{K}_{t}$, is a linear, infinite-dimensional operator that acts on $g$ in the following manner, $\mathcal{K}_{t}g := g \circ \bm{S}_{t}$, where:

\vspace{-.7cm}\begin{equation}
\bm{S}_{t}:\mathbb{X}\to\mathbb{X};\,x(0)\to x(t)=x(0)+\int_{0}^{t}\bm{f}(\bm{x}(\tau))d\tau    
\end{equation}

\vspace{-.1cm}\noindent
is called the \emph{flow}. The \emph{Koopman eigenvalues}, $\lambda$, and \emph{Koopman eigenfunctions}, $\phi(\bm{x})$, of (\ref{eq.1z}) are such that:

\vspace{-.2cm}\begin{equation}
\mathcal{K}_{t}\phi_{i}=e^{\lambda_{i}t}\phi_{i}, \quad i=1,...,\infty,
\end{equation}

\vspace{-.1cm}\noindent
where $\lambda_{i}\in\mathbb{C}$, and $\phi_{i}\in\mathcal{F}$ is nonzero. Now, consider a vector-valued function, $\bm{g}:\mathbb{X}\to\mathbb{C}^{q}$. If all elements of $\bm{g}$ lie within the span of the Koopman eigenfunctions, then:

\vspace{-.4cm}\begin{equation}\label{eq.5x}
\bm{g}(\bm{x}(t)) = \sum_{i=1}^{\infty}\phi_{i}(\bm{x}(t))\bm{\upsilon}_{i} = \sum_{i=1}^{\infty}\phi_{i}(\bm{x}(0))\bm{\upsilon}_{i}e^{\lambda_{i}t},
\end{equation}

\vspace{-.1cm}\noindent
where $\bm{\upsilon}_{i}\in\mathbb{C}$ are referred to as \emph{Koopman modes} \cite{Mauroy2020}.

\vspace{-.2cm}
\vspace{-.1cm}\subsection{Extended Dynamic Mode Decomposition \cite{Williams2015}}
Consider pairs of snapshots of the system state variables, $\{\bm{x}_{k-1},\bm{x}_{k}\}$, $k=1,...,m$, as sampled data of continuous flows, i.e., under a sampling period $\Delta t$, we have $x_{k}=x(k\Delta t)$. The data matrices are defined as $\bm{X}_{0}=[\bm{x}_{0}\, ...\, \bm{x}_{m-1}]$, $\bm{X}_{1}=[\bm{x}_{1}\, ...\, \bm{x}_{m}]$, where $\bm{X}_{0},\bm{X}_{1}\in\mathbb{R}^{n\times m}$. The vector of observable functions is defined as $\bm{g}\left(\bm{x}_{k}\right):=\left[g_{1}\left(\bm{x}_{k}\right)\, ...\, g_{q}\left(\bm{x}_{k}\right)\right]\tran$, where $\bm{g}:\mathbb{R}^{n} \to \mathbb{R}^{q}$, $q \ge n$. Also, the matrices of observables are defined as $\bm{G}_{0}=\left[\bm{g}\left(\bm{x}_{0}\right)\, ...\, \bm{g}\left(\bm{x}_{m-1}\right)\right]$, $\bm{G}_{1}=\left[\bm{g}\left(\bm{x}_{1}\right)\, ...\, \bm{g}\left(\bm{x}_{m}\right)\right]$, where $\bm{G}_{0},\bm{G}_{1}\in\mathbb{R}^{q\times m}$. A finite-dimensional approximation to the Koopman operator is estimated as follows:

\vspace{-.5cm}\begin{equation}\label{eqkoopmanoperatorestimation}
\bm{K}=\bm{G}_{1}\bm{G}_{0}^{\dagger},
\end{equation}

\vspace{-.2cm}\noindent
where $\bm{G}_{0}^{\dagger}$ denotes the Moore-Penrose pseudoinverse of $\bm{G}_{0}$, and $\bm{K}\in\mathbb{R}^{q\times q}$. The eigenvalues $\mu \approx e^{(\lambda\cdot\Delta t)}$ of $\bm{K}$ provide an approximation to the Koopman eigenvalues, whereas an approximation to the Koopman eigenfunctions is given by:

\vspace{-.2cm}\begin{equation}\label{eqeigenfunction}
\bm{\phi}\left(\bm{x}_{k}\right)\approx\bm{L}\bm{g}\left(\bm{x}_{k}\right),
\end{equation}

\vspace{-.1cm}\noindent
where $\bm{L}$ contains the left eigenvectors of $\bm{K}$, and $\bm{\phi}\left(\bm{x}_{k}\right)=\left[\phi_{1}\left(\bm{x}_{k}\right)\, ...\, \phi_{q}\left(\bm{x}_{k}\right)\right]\tran$. Finally, to recover the Koopman modes for the full set of state variables $\bm{g}\left(\bm{x}_{k}\right)=\bm{x}_{k}$, let the projection matrix, $\bm{P}\in\mathbb{R}^{n\times q}$, be a matrix defined such that $\bm{x}_{k}=\bm{P}\bm{g}\left(\bm{x}_{k}\right)$. From (\ref{eqeigenfunction}), we have that $\bm{g}\left(\bm{x}_{k}\right)=\bm{L}^{-1}\bm{\phi}\left(\bm{x}_{k}\right)$, and thus, $\bm{x}_{k}=\bm{P}\bm{g}\left(\bm{x}_{k}\right)=\bm{P}\bm{L}^{-1}\bm{\phi}\left(\bm{x}_{k}\right)$. Hence, an approximation to the Koopman modes is provided by the column vectors of $\bm{U}=\bm{P}\bm{L}^{-1}$, $\bm{U}\in\mathbb{C}^{n\times q}$, and:

\vspace{-.3cm}\begin{equation}\label{eqedmdexpansion}
\bm{x}_{k} \approx \sum_{i=1}^{q}\phi_{i}\left(\bm{x}_{k}\right)\bm{\upsilon}_{i}=\sum_{i=1}^{q}\phi_{i}\left(\bm{x}_{0}\right)\bm{\upsilon}_{i}\mu_{i}^{k}.
\end{equation}

\vspace{-.1cm}
Eq. (\ref{eqedmdexpansion}) is a finite truncation of (\ref{eq.5x}) under the sampling.

\vspace{-.1cm}
\section{Construction of Observable Functions}\label{sec.III}

This section contains the main contribution of this letter. Let $\mathbb{X}\subseteq\mathbb{R}^{n}$ in the continuous time system (\ref{eq.1z}). In what follows, we drop the time index, $t$, for simplicity. We are interested in the case where the nonlinear functions, $\bm{f}(\bm{x})$, can be written as a linear combination of elementary functions, $h(\bm{x})\in\mathcal{F}$---that is, if we consider the $i-$th state variable:

\vspace{-.3cm}\begin{equation}
\dot{x}_{i} = c_{00} + \bm{c}_{0}\tran\bm{x} + c_{1}h_{1}(\bm{x}) + \cdots + c_{r}h_{r}(\bm{x}), \label{eq.12}
\end{equation}

\vspace{-.2cm}\noindent
where $\bm{c}_{0}\tran$ denotes the transpose of $\bm{c}_{0}$. The elementary functions include $\sin{x}$, $\cos{x}$, $e^{x}$, and $\frac{x}{b+x}$, as well as compositions of these elementary functions. Note that because of the composition of functions, these elementary functions encompass a broad class of models encountered in engineering, making this approach well-suited to applications of KOT. Indeed, mathematical models of many engineering systems can be written in the form of (\ref{eq.12})---including models of ion channels \cite{Sigg2014}, semiconductor devices \cite{Grasser2003}, and power systems \cite{Netto2018b, Netto2018c}---thereby motivating the search for a state-inclusive Koopman observable space \cite{Johnson2018}. The following procedure was proposed in \cite{Gu2011} as part of a model order reduction method, and it was recently applied in the context of system identification on a lifted space \cite{Qian2020}. For each elementary function, $h_{i}(\bm{x})$,

\begin{enumerate}
\item Introduce a new variable $z_{i}=h_{i}(\bm{x})$.
\item Replace $h_{i}(\bm{x})$ by $z_{i}$ in the original equations.
\item Add $\dot{z}_{i}=\frac{\partial h_{i}(\bm{x})}{\partial\bm{x}}\bm{f}$ in the set of original equations.
\end{enumerate}

The equation added in Step $3$ is the Lie derivative of $z_{i}$ with respect to $\bm{f}$. Note that the Lie derivative is a Koopman generator in terms of the vector field $\bm{f}$. The resulting lifted system is as follows:

\vspace{-.5cm}\begin{align}
\dot{x}_{i} &= c_{00} + \bm{c}_{0}\tran\bm{x} + c_{1}z_{1} + \cdots + c_{r}z_{r}, \\
\dot{z}_{i} &= \mathcal{L}_{\bm{f}}h_{i}(\bm{x}),
\end{align}

\vspace{-.2cm}\noindent
where $\mathcal{L}_{\bm{f}}h_{i}(\bm{x})=\frac{\partial h_{i}(\bm{x})}{\partial x_{1}}\dot{x}_{1}+\cdots+\frac{\partial h_{i}(\bm{x})}{\partial x_{n}}\dot{x}_{n}$. Table \ref{tab.univar} shows examples of transformations for univariate elementary functions, and the following remarks are in order:

\begin{enumerate}
\item $x^{-1}$ can be removed from the new differential equations by introducing another new variable, $y=x^{-1}$.
\item There are elementary functions that need to be handled by adding two new variables, e.g., $\sin{x}$.
\end{enumerate}

Now, for compositions of elementary functions, i.e., $h(\bm{x})=(h_{1}\circ h_{2})(\bm{x})=h_{1}(h_{2}(\bm{x}))$, proceed as follows:

\begin{enumerate}
\item Introduce new variables $z_{1}=h_{1}(\bm{x})$ and $z_{2}=h_{2}(z_{1})$.
\item Replace $h_{2}(h_{1}(\bm{x}))$ by $z_{2}$ in the original equations.
\item Add $\dot{z}_{1}=\frac{\partial h_{1}(\bm{x})}{\partial\bm{x}}\bm{f}$ and $\dot{z}_{2}=\frac{\partial h_{2}(z_{1})}{\partial z_{1}}\dot{z}_{1}$ in the set of original equations.
\end{enumerate}

\begin{table}[!t]
\centering
\setlength{\tabcolsep}{0.5em}
\caption{Transformations for univariate elementary functions}
\vspace{-.25cm}
\begin{tabular}{l l l}
\hline
Elementary function  & New variable(s)   & New differential equation(s) \\ \hline
$h(x)=e^{x}$         & $z=e^{x}$         & $\dot{z}=e^{x}\dot{x}=z\dot{x}$ \\ \hline
$h(x)=\frac{1}{b+x}$ & $z=\frac{1}{b+x}$ & $\dot{z}=-\frac{1}{(b+x)^{2}}\dot{x}=-z^{2}\dot{x}$ \\ \hline
$h(x)=\ln{x}$        & $z_{1}=\ln{x}$    & $\dot{z}=x^{-1}\dot{x}=z_{2}\dot{x}$ \\ 
                     & $z_{2}=x^{-1}$    & $\dot{z}_{2}=-x^{-2}\dot{x}=-z_{2}^{2}\dot{x}$ \\ \hline
$h(x)=\sin{x}$       & $z_{1}=\sin{x}$   & $\dot{z}_{1}=(\cos{x})\dot{x}=z_{2}\dot{x}$ \\
                     & $z_{2}=\cos{x}$   & $\dot{z}_{2}=(-\sin{x})\dot{x}=-z_{1}\dot{x}$ \\
\hline
\end{tabular}
\label{tab.univar}
\end{table}

\vspace{-.4cm}\noindent
\begin{table}[!t]
\centering
\setlength{\tabcolsep}{0.2em}
\caption{Examples of polynomialization of systems given by composition of elementary functions}
\vspace{-.25cm}
\begin{tabular}{l l l}
\hline
Original system                           & New variables               & Lifted system \\ \hline
$\dot{x}=\frac{1}{1+e^{-x}}$ &                             & $\dot{x}=z_{2}$ \\ 
                                          & $z_{1}=e^{-x}$ & $\dot{z}_{1}=-e^{-x}\frac{1}{1+e^{-x}}=-z_{1}z_{2}$ \\ 
                                          & $z_{2}=\frac{1}{1+z_{1}}$   & $\dot{z}_{2}=-\frac{1}{(1+z_{1})^{2}}(-z_{1}z_{2})=z_{1}z_{2}^{3}$ \\ \hline
$\dot{x}=x\cos{x}$                        &                             & $\dot{x}=z_{2}$ \\
                                          & $z_{1}=\cos{x}$             & $\dot{z}_{1}=-\sin{x}(x\cos{x})=-z_{2}z_{3}=-z_{1}z_{4}$ \\
                                          & $z_{2}=x z_{1}$             & $\dot{z}_{2}=x\cos{x}\cos{x}-x^{2}\sin{x}\cos{x}$ \\
                                          &                             & $\quad\;=z_{1}z_{2}-z_{2}z_{4}$ \\
                                          & $z_{3}=\sin{x}$             & $\dot{z}_{3}=\cos{x}(x\cos{x})=z_{1}z_{2}$ \\
                                          & $z_{4}=x z_{3}$             & $\dot{z}_{4}=x\cos{x}\sin{x}+x^{2}\cos{x}\cos{x}$ \\
                                          &                             & $\quad\;=z_{2}z_{3}+z_{2}^{2}=z_{1}z_{4}+z_{2}^{2}$ \\
\hline
\end{tabular}
\label{tab.comp}
\end{table}

Two examples of polynomialization involving the composition of elementary functions are given in Table \ref{tab.comp}, from which two remarks are in order:
\begin{enumerate}
\item Elementary functions that need to be handled by adding two new variables must be considered.
\item Polynomialization is not a unique transformation, as is clear from the second example.
\end{enumerate}

The polynomial system resulting from the previous procedure often contains terms of third and higher order; these terms can always be eliminated by defining an additional set of observable functions and by subsequently applying the lifting procedure once more. Finally, a quadratic-linear system can be obtained. Upon completion of the polynomialization, we select the state variables, along with the obtained new variables, as observable functions in the EDMD, i.e., $\{x_{1},...,x_{n},z_{1},z_{2},...\}$. This choice of observable functions is justified by the fact that polynomialization is an exact transformation, from the original state space to a higher dimension space; therefore, the lifted representation serves as a weak canonical form of the original system, given its nonuniqueness.

\vspace{-.1cm}
\section{Numerical Results}\label{sec.IV}
KOT is gaining momentum in the power system community \cite{Susuki2016}. Power systems are chosen as test systems in this work because they have multiple attractors, and their model contains sine and cosine functions. Consider:

\vspace{-.5cm}
\begin{equation}
\dot{\delta} = \omega, \qquad
\dot{\omega} = c_{00} + c_{0}\omega + c_{1}\cos\delta + c_{2}\sin\delta. \label{eq.x1}
\end{equation}

\begin{figure}[!t]
\centering
\begin{minipage}[b]{.255\textwidth}\hspace{-.5cm}
\centering\subfloat[][\label{fig.phase_portrait}{}]{
\includegraphics[width=1\columnwidth]{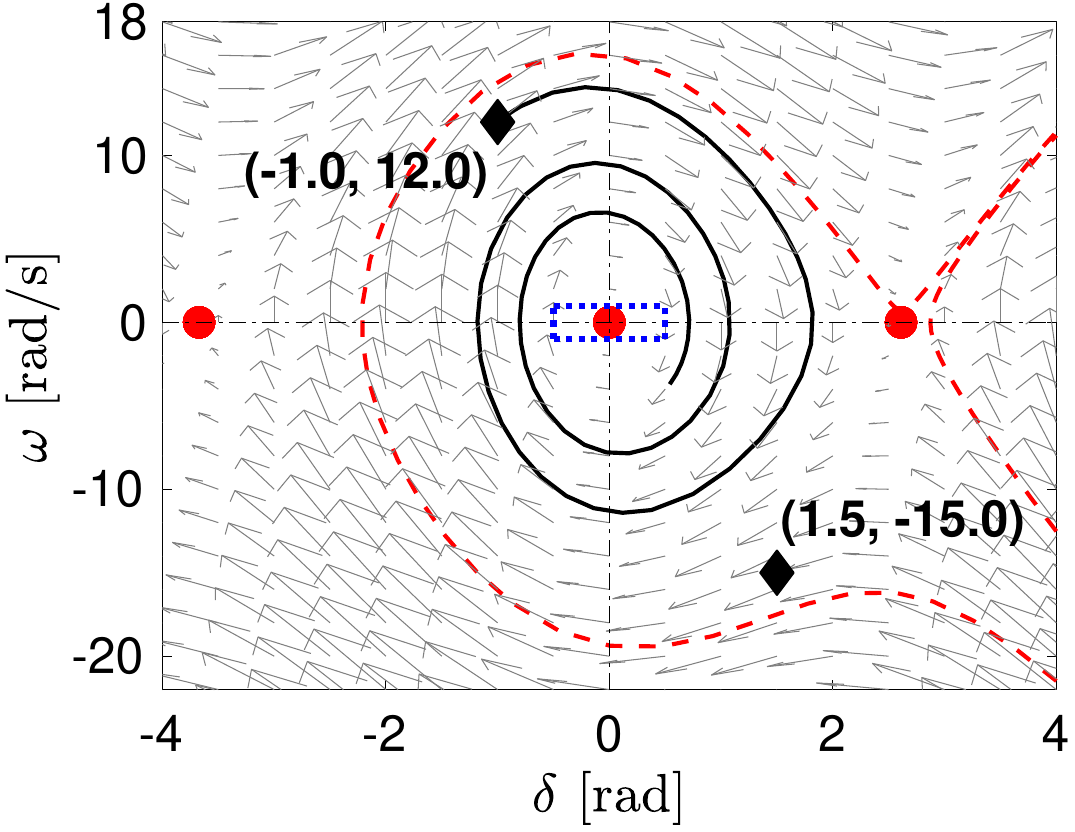}}
\end{minipage}
\begin{minipage}[b]{.255\textwidth}\hspace{-.5cm}
\centering\subfloat[][\label{fig.linear}{}]{
\includegraphics[width=1\columnwidth]{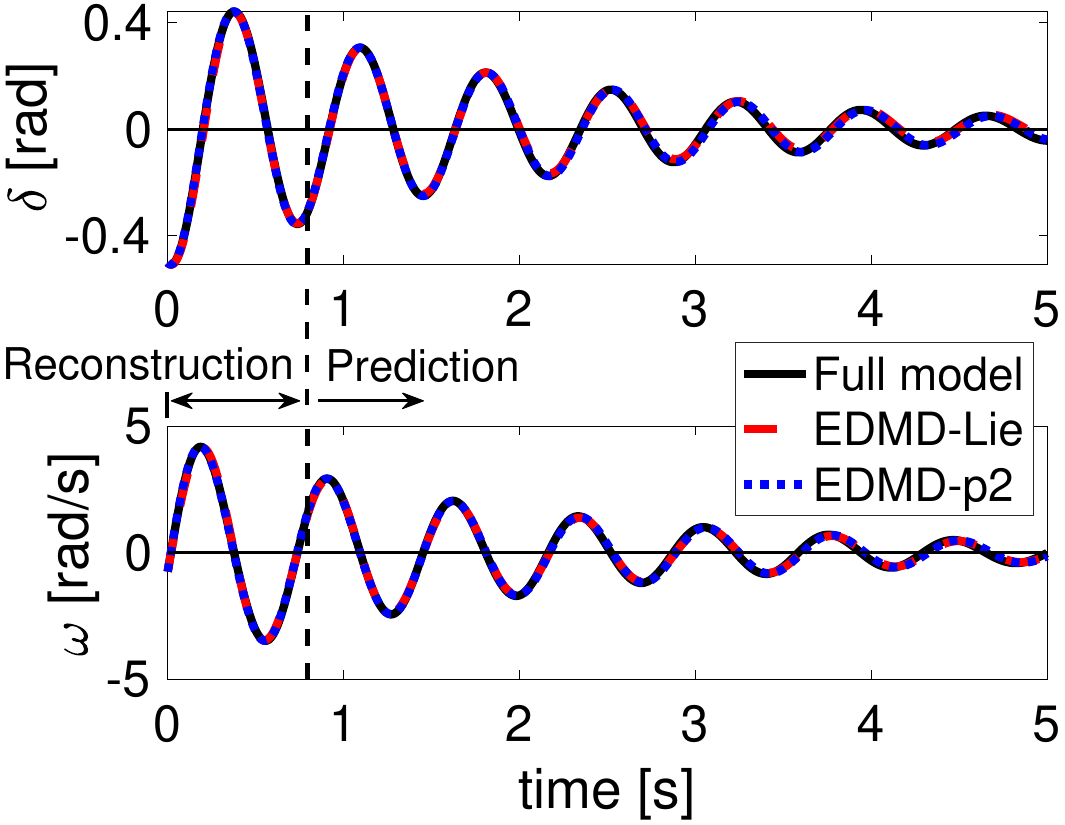}}
\end{minipage}
\vspace{-.4cm}
\caption{(a) Phase portrait of (\ref{eq.x1}). The symbol \tikzcircle{2pt} denotes fixed points, and $\MyDiamond[draw=black,fill=black]$ denotes initial states $\left(\delta_{0},\omega_{0}\right)$ for cases $1$--$2$ in Fig. \ref{fig.nonlinear_mild}. The dashed line {\color{red}\hdashrule[0.4ex]{0.6cm}{0.7pt}{0.8mm}} delineates the attractor. The dotted line {\color{blue}\hdashrule[0.4ex]{0.5cm}{1pt}{1pt}} rectangle centered at the origin indicates the region where trajectories were sampled to compute the EDMD. (b) Example of a trajectory starting at $\left(\delta_{0},\omega_{0}\right)=\left(-0.50,-0.75\right)$, i.e., within the sampled region surrounding the fixed point $(0,0)$ where trajectories were sampled to compute the EDMD. The system mode is mildly damped, $\mu_{\text{true}}=-0.5000\pm j8.8503$, $f_{\text{true}}=1.4086$Hz, and $\xi_{\text{true}}=5.6406$\%. }
\label{fig.xxx}
\end{figure}

\vspace{-.1cm}
Details of the power system model (\ref{eq.x1}) are provided in Appendix \ref{sec.appA}, and its phase portrait is shown in Fig. \ref{fig.phase_portrait}. Trajectories with starting point $(\delta_{0},\omega_{0})$ in the lattice $\delta=(-0.50:0.25:0.50)$, $\omega=(-1.00:0.25:1.00)$, are sampled at each $\Delta t=0.005$ second. This sampling rate is consistent with the available technology for power system measurement devices---namely, phasor measurement units. The lattice is indicated in Fig. \ref{fig.phase_portrait} by the blue rectangle centered at the origin, and it contains $45$ starting points, thereby leading to $45$ sampled trajectories. Note that all the trajectories in the lattice are in a linear region of the state space. We record the initial $0.8$ second of each trajectory, thereby leading to $160$ samples per trajectory. These trajectories are used to compute EDMD.


Now, define ${z}_{1}:=\delta$, ${z}_{2}:=\omega$, ${z}_{3}:=\sin\delta$, and ${z}_{4}:=\cos\delta$. By applying the procedure outlined in Section \ref{sec.III}, we have:

\vspace{-.5cm}\begin{align}
\dot{z}_{1} &= {z}_{2}, \qquad\qquad\quad\;\,\,\,
\dot{z}_{2} = c_{00} + c_{0}{z}_{2} + c_{1}{z}_{4} + c_{2}{z}_{3}, \nonumber \\
\dot{z}_{3} &= \mathcal{L}_{\bm{f}}\sin\delta = {z}_{2}{z}_{4}, \;
\dot{z}_{4} = \mathcal{L}_{\bm{f}}\cos\delta = -{z}_{2}{z}_{3}, \label{eq4.embedding}
\end{align}

\vspace{-.1cm}\noindent
and (\ref{eq4.embedding}) contains only monomials in $z$. The lifted dynamical system in $z$ suggests the use of the following observable functions, $\left[ {z}_{1}\;\; {z}_{2}\;\; {z}_{3}\;\; {z}_{4}\;\; {z}_{2}{z}_{3}\;\; {z}_{2}{z}_{4} \right]\tran$, which yields:

\vspace{-.2cm}\begin{equation}
\bm{g} = \left[ \delta\;\; \omega\;\; \sin\delta\;\; \cos\delta\;\; \omega\sin\delta\;\; \omega\cos\delta \right]\tran. \label{eq.observables}
\end{equation}

\vspace{-.1cm}
We compute EDMD using the observable functions given by (\ref{eq.observables}), which we refer to as EDMD-Lie. Additionally, for comparison, we compute EDMD with other sets of observable functions that have been widely used in the literature. In what follows, EDMD-p\underline{N} denotes the case where all monomials of the state variables up to degree \underline{N} are used. For example, in the case of EDMD-p3, $\bm{g}=\left[\delta\; \omega\; \delta\omega\; \delta^{2}\; \omega^{2}\; \delta^{2}\omega\; \delta\omega^{2}\; \delta^{3}\; \omega^{3}\right]\tran$. We consider using all monomials of the state variables up to degrees $2$, $3$, and $4$, respectively denoted by EDMD-p2, EDMD-p3, and EDMD-p4. Further, the use of radial basis functions is also common in the literature. EDMD-rbf\underline{N} denotes the case where the state variables, $\left(\delta,\omega\right)$, plus \underline{N}$-2$ thin-plate spline radial basis functions with center at $\bm{x}_{c}$ are used \cite{Korda2018b}; hence, $g\left(\bm{x}\right) = \norm{\bm{x}-\bm{x}_{c}}^{2}\log\left(\norm{\bm{x}-\bm{x}_{c}}\right)$. We consider two cases with radial basis functions. The first case, EDMD-rbf6, is designed to have the same size of the Lie lifted system. This will allow us to make a fair comparison between EDMD-Lie and the case where radial basis functions are used. The second case, EDMD-rbf19, is designed to exploit the maximum possible number of radial basis functions before the Koopman Kalman filter (KKF) \cite{Netto2018a, Netto2018b} in Section \ref{sec.KOF} becomes unobservable. As mentioned in Section \ref{sec.KOF}, the KKF relies on measurements of real and reactive power acquired at the generator terminal. For the single-machine infinite-bus system, the rank of the measurement matrix of the KKF turns deficient beyond the 19 canonical variables defined by the radial basis functions.

We define three criteria to assess the performance of the EDMD in approximating the Koopman operator:

\begin{enumerate}
\item \emph{Accuracy in the estimation of principal eigenvalues.}
\item \emph{Accuracy in the prediction of unknown trajectories.}
\item \emph{Number of observable functions.}
\end{enumerate}

These criteria are used in what follows. By linearizing (\ref{eq.x1}) around the fixed point, $(0,0)$, and by computing the eigenvalues of the obtained Jacobian matrix, one finds a pair of complex-conjugate eigenvalues, $\mu_{\text{true}}=-0.5000\pm j8.8503$, associated with the linear mode of frequency, $f_{\text{true}}=1.4086$~Hz, and damping ratio, $\xi_{\text{true}}=5.6406$~\%. For comparison, the eigenvalues estimated through EDMD with different sets of observable functions are shown in Table \ref{tab.evals}. The pairs of complex-conjugate eigenvalues associated with the linear mode, referred to as principal eigenvalues, are shaded in blue. Note that the principal eigenvalues are well approximated in all cases; however, although estimating the linear mode with good numerical accuracy is important, this is only part of the information needed to represent the entire domain of attraction through the approximated Koopman tuples, $\{\lambda,\phi,\upsilon\}$.



\noindent

\begin{table}[!t]
\centering \scriptsize
\setlength{\tabcolsep}{0.1em}
\caption{Eigenvalues estimated through EDMD with different sets of observable functions (principal eigenvalues are shaded)}
\vspace{-.25cm}
\begin{tabular}{l | r r r}
\hline
\hspace{.1em} EDMD \hspace{.1em} & \multicolumn{3}{c}{Eigenvalues} \\ \hline
\hspace{.1em} Lie   & $ 0.0529              $ & \cellcolor{NREL1}$-0.5134 \pm j  8.7623$ & $-3.5394 \pm j 17.7558$ \\
                    & $-5.4014              $ &                         &                                          \\ \hline
\hspace{.1em} p2    & \cellcolor{NREL1}$-0.5011 \pm j  8.7504$ & $-0.9950 \pm j 17.4716$ & $-1.0048              $ \\ \hline
\hspace{.1em} p3    & \cellcolor{NREL1}$-0.4943 \pm j  8.8511$ & $-0.9961 \pm j 17.4634$ & $-1.0095              $ \\
                    & $-1.4837 \pm j 26.1940$ & $-1.4938 \pm j  8.6778$ &                                          \\ \hline
\hspace{.1em} p4    & \cellcolor{NREL1}$-0.4949 \pm j  8.8487$ & $-0.9734 \pm j 17.6953$ & $-1.0001              $ \\
                    & $-1.4782 \pm j 26.1682$ & $-1.5271 \pm j  8.6792$ & $-1.9806 \pm j 34.9201$                  \\
                    & $-1.9844              $ & $-2.0156 \pm j 17.3382$ &                                          \\ \hline
\hspace{.1em} rbf6  & $-0.2218 \pm j  1.3005$ & \cellcolor{NREL1}$-0.5059 \pm j  8.7427$ & $-1.2829 \pm j 14.2659$ \\ \hline
\hspace{.1em} rbf19 & $-0.0463              $ & \cellcolor{NREL1}$-0.4976 \pm j  8.7607$ & $-1.2354 \pm j 19.2201$ \\
                    & $-2.4783 \pm j  1.8160$ & $-6.9312 \pm j 11.6575$ & $-7.7604 \pm j 26.8570$                  \\
                    & $-9.1463 \pm j 20.7501$ & $-12.7618\pm j 51.9406$ & $-16.5766\pm j 93.1681$                  \\
                    & $-31.6343\pm j 42.8542$ &                         &                                          \\
\hline
\end{tabular}
\label{tab.evals}
\end{table}

\noindent
\begin{figure}[!t]
\centering
\begin{minipage}[b]{.255\textwidth}\hspace{-.7cm}
\centering\subfloat[][\label{kkf2w}{Case $1$: $\left(\delta_{0},\omega_{0}\right)=\left(-1.0,12.0\right)$}]{
\includegraphics[width=.96\columnwidth]{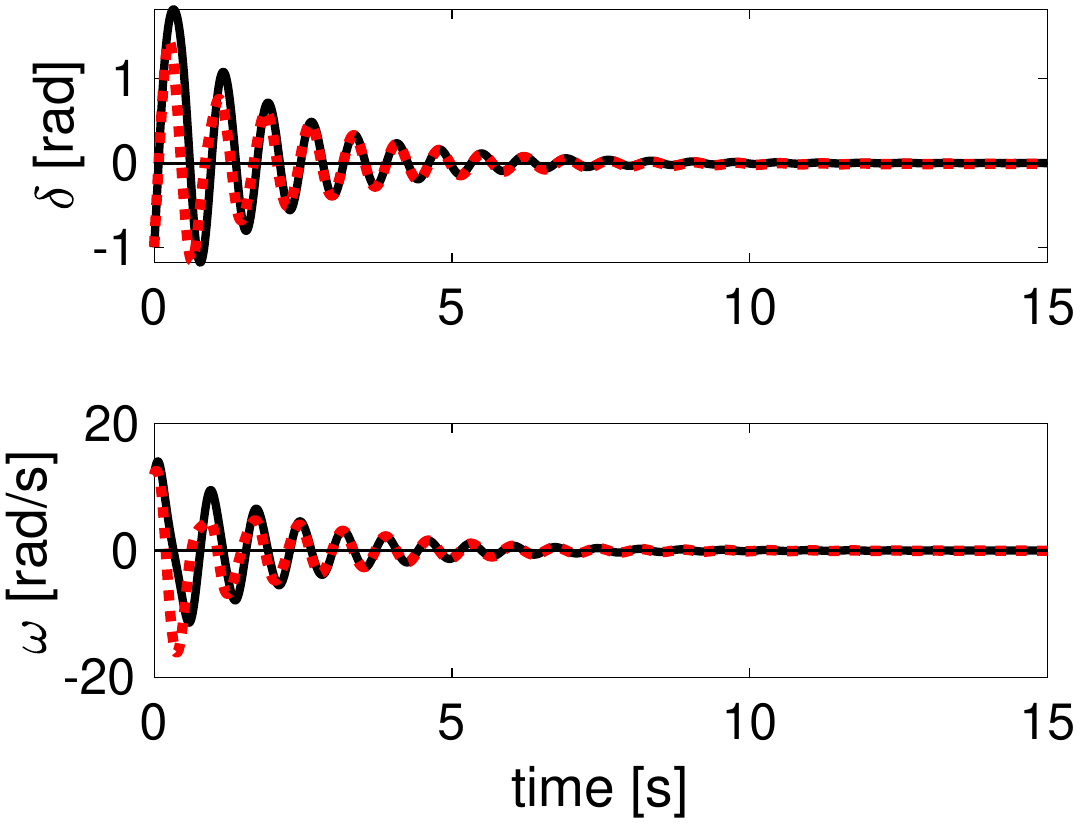}}
\end{minipage}
\begin{minipage}[b]{.255\textwidth}\hspace{-.7cm}
\centering\subfloat[][\label{kkf3w}{Case $2$: $\left(\delta_{0},\omega_{0}\right)=\left(1.5,-15.0\right)$}]{
\includegraphics[width=.96\columnwidth]{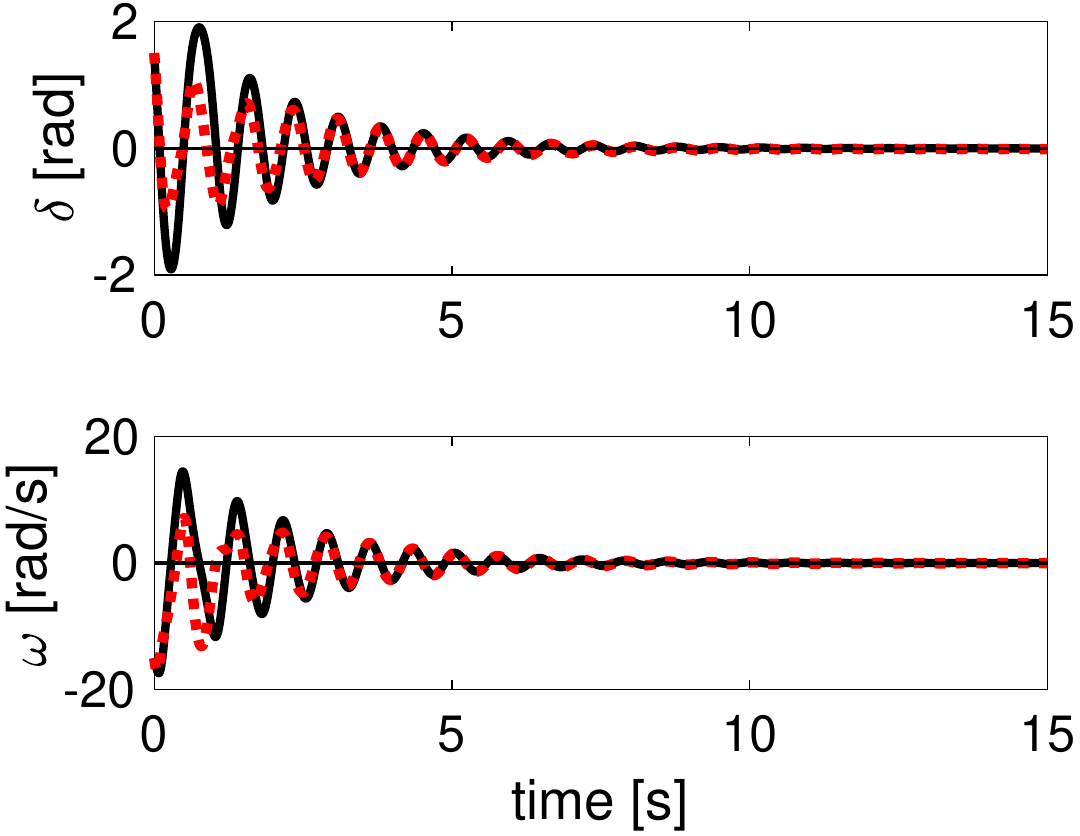}}
\end{minipage}\\
\begin{minipage}[b]{.255\textwidth}\hspace{-.7cm}
\centering\subfloat[][\label{conv2}{Case $1$: Absolute error in $\delta$}]{
\includegraphics[width=.96\columnwidth]{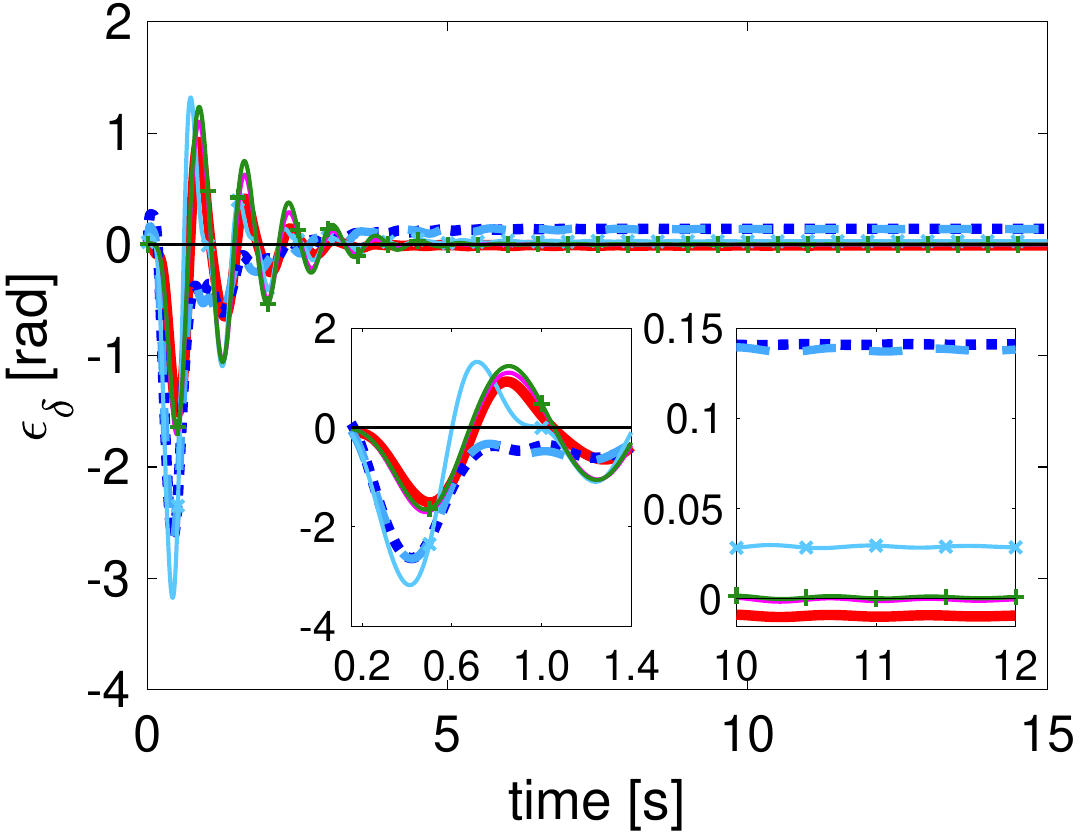}}
\end{minipage}
\begin{minipage}[b]{.255\textwidth}\hspace{-.7cm}
\centering\subfloat[][\label{conv3}{Case $2$: Absolute error in $\delta$}]{
\includegraphics[width=.96\columnwidth]{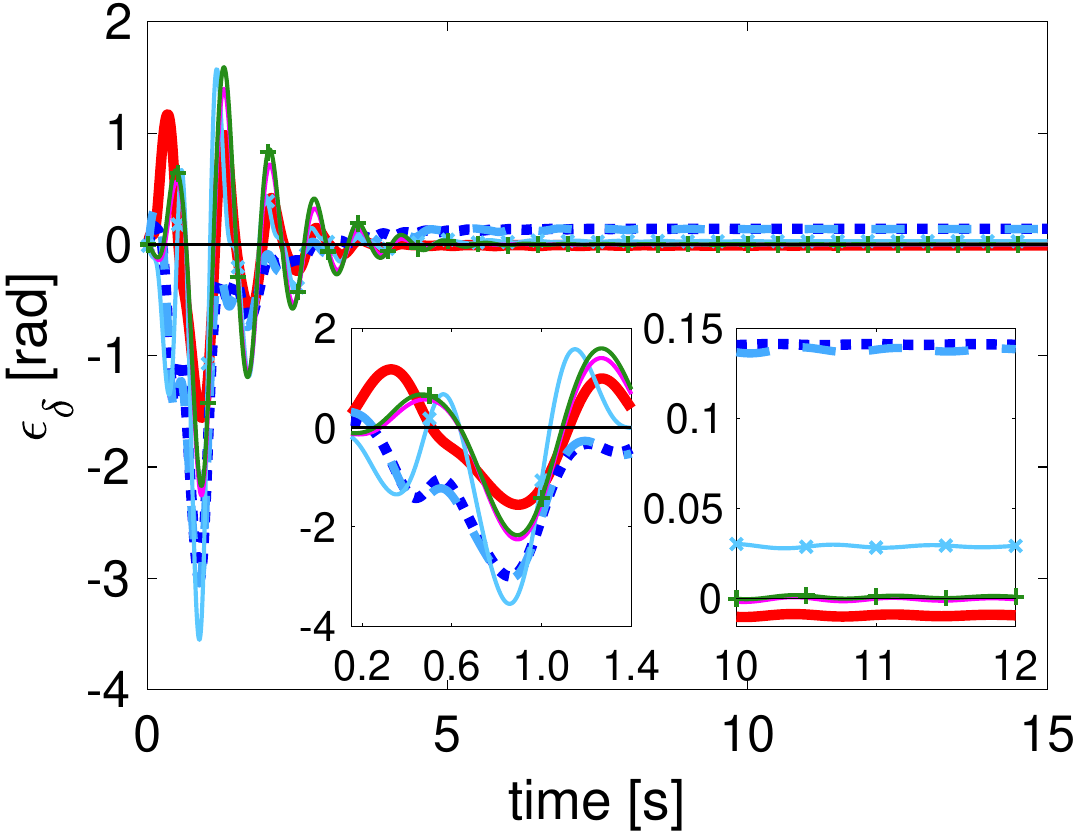}}
\end{minipage}
\vspace{-.3cm}
\caption{Example of trajectories starting near the stability boundary.}
\label{fig.nonlinear_mild}
\end{figure}

\begin{table}[!t]
\centering \scriptsize
\setlength{\tabcolsep}{0.4em}
\caption{Statistics of the errors in $\delta$ and $\omega$ associated with cases $1$--$2$ shown in Fig. \ref{fig.nonlinear_mild} (the least error is shaded)}
\vspace{-.25cm}
\begin{tabular}{l l r r r r r r}
\hline
 & & \multicolumn{6}{c}{EDMD} \\ \cline{3-8}
Case & Statistics & Lie & p2 & p3 & p4 & rbf6 & rbf19 \\ \hline
$1$ & $\max{\epsilon_{\delta}}$     & \cellcolor{NREL1}$ 1.5096$ & $ 2.6313$ & $ 2.6418$ & $ 3.1758$ & $ 1.7058$ & $ 1.6560$ \\
    & $\max{\epsilon_{\omega}}$     & $14.6195$ & $23.8467$ & $21.9599$ & $32.1051$ & \cellcolor{NREL1}$12.8223$ & $12.9307$ \\
    & $\sum{\epsilon_{\delta}}/1e2$ & \cellcolor{NREL1}$ 2.2946$ & $ 5.9485$ & $ 5.9178$ & $ 3.8145$ & $ 2.9216$ & $ 3.1676$ \\
    & $\sum{\epsilon_{\omega}}/1e3$ & \cellcolor{NREL1}$ 2.1780$ & $ 6.4989$ & $ 5.4375$ & $ 4.0880$ & $ 2.3849$ & $ 2.6352$ \\
\hline
$2$ & $\max{\epsilon_{\delta}}$     & \cellcolor{NREL1}$ 1.5572$ & $ 3.0010$ & $ 3.0463$ & $ 3.5522$ & $ 2.2549$ & $ 2.1697$ \\
    & $\max{\epsilon_{\omega}}$     & \cellcolor{NREL1}$12.7739$ & $29.5142$ & $26.7154$ & $36.9509$ & $16.5962$ & $16.7258$ \\
    & $\sum{\epsilon_{\delta}}/1e2$ & \cellcolor{NREL1}$ 3.0947$ & $ 6.9762$ & $ 7.1406$ & $ 4.7190$ & $ 3.8288$ & $ 4.1487$ \\
    & $\sum{\epsilon_{\omega}}/1e3$ & \cellcolor{NREL1}$ 2.6237$ & $ 8.2182$ & $ 6.9028$ & $ 5.8950$ & $ 3.1748$ & $ 3.4745$ \\
\hline
\end{tabular}
\label{tab.errors}
\end{table}

\begin{figure}[!t]
\centering
\includegraphics[width=0.8\columnwidth]{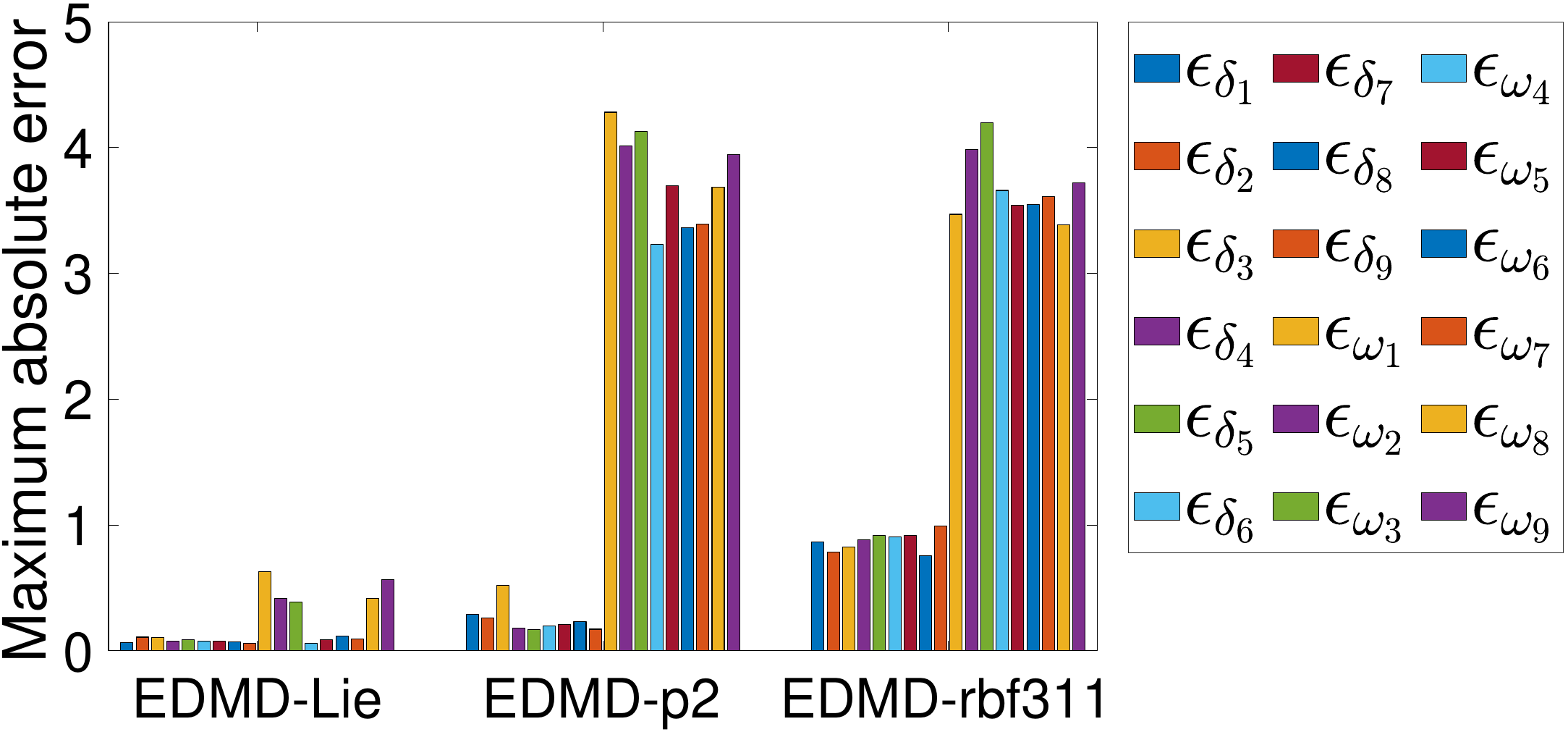}
\vspace{-.4cm}
\caption{Maximum absolute error in the prediction of a strongly nonlinear trajectory for each of the 18 system state variables. The error in $\delta$ is in [rad], while the error in $\omega$ is in [rad/s].}
\label{fig.multimachine}
\end{figure}

\vspace{-1cm}\subsection{Reconstruction and Prediction of Known Trajectories for a Single-Machine Infinite-Bus System}
To further assess the performance of the EDMD with different sets of observable functions, we use the approximated Koopman tuples to reconstruct known trajectories. By \emph{known trajectories}, we mean trajectories given as inputs to the EDMD algorithm. In Fig. \ref{fig.linear}, we show the results obtained with EDMD-Lie and EDMD-p2. We omit the results obtained with other sets of observable functions because they are quantitatively and qualitatively similar to EDMD-Lie. We observe that these trajectories do not pose any challenge to the EDMD, independent of the choice of observable functions, because they are in a linear region of the state space. In this case, however, it is difficult to assess whether the EDMD is performing well or simply overfitting the a priori known input data. 

\vspace{-.3cm}\subsection{Prediction of Strongly Nonlinear, Unknown Trajectories for a Single-Machine Infinite-Bus System}\label{sec.KOF}

In theory, the Koopman operator is valid in the entire domain of attraction; hence, the approximation of the Koopman operator via EDMD should provide the means to predict, with good numerical accuracy, any trajectory in the same domain of attraction for which the EDMD was computed. This is when the choice of observable functions plays a key role---for example, it will directly affect the transient stability analysis of electric power grids. We use the approximated Koopman tuples to predict trajectories that start far from the fixed point and are strongly nonlinear. Further, these trajectories are not used as inputs to the EDMD in the first place. This test will reveal how well the Koopman tuples approximated by the EDMD are representative of the entire domain of attraction. Four trajectories are selected, with initial states indicated by $\MyDiamond[draw=black,fill=black]$ in Fig. \ref{fig.phase_portrait}. The Koopman tuples estimated through the EDMD are used to design a robust KKF \cite{Netto2018a, Netto2018b}, with real and reactive power measured at the generator terminal. Figs. \ref{kkf2w}--\ref{kkf3w} compare the trajectory obtained with the KKF design with EDMD-Lie versus the full nonlinear model, (\ref{eq.x1}), for each selected starting point. To avoid making the plot too crowded, Figs. \ref{kkf2w}--\ref{kkf3w} do not show the trajectories obtained with the KKF design using other sets of observable functions; instead, we calculate the absolute error in $\delta$ and $\omega$ for each case and present their statistics in Table \ref{tab.errors}. Also, the absolute error in $\delta$ is shown in Figs. \ref{conv2}--\ref{conv3}. Table \ref{tab.evals} shows that EDMD-Lie is less accurate in the estimation of principal eigenvalues. On the other hand, Table \ref{tab.errors} and Fig. \ref{fig.nonlinear_mild} show that EDMD-Lie has overall better performance, indicating that the proposed analytical approach to ascertain observable functions captures the system dynamics well enough to predict unforeseen scenarios. In this illustration, the availability of the system model enabled us to make an informed decision on the selection of observable functions. Further, EDMD-Lie yields less observable functions than EDMD-p3, EDMD-p4, and EDMD-p19, thereby providing a computational advantage.

\vspace{-.3cm}
\subsection{Prediction of Strongly Nonlinear, Unknown Trajectories for a Multimachine Power System}
Now, consider the New England 39-bus test system \cite{Susuki2011}. Following the proposed analytical construction, the observable functions are given by:

\vspace{-.1cm}\noindent
\begin{align}
    \bm{g} = [\delta_{\ell}&\;\; \omega_{\ell}\;\; \omega_{\ell}\sin\delta_{\ell}\;\; \omega_{\ell}\cos\delta_{\ell}\;\; \cos\delta_{\ell}\cos\delta_{p}\;\; \sin\delta_{\ell}\sin\delta_{p} \nonumber\\
    &\sin\delta_{\ell}\cos\delta_{p}\;\; \cos\delta_{\ell}\sin\delta_{p}\;\; \sin\delta_{\ell}\cos\delta_{p}\omega_{\ell} \nonumber\\ 
    &\;\;\cos\delta_{\ell}\sin\delta_{p}\omega_{\ell}\;\; \cos\delta_{\ell}\cos\delta_{p}\omega_{\ell}\;\; \sin\delta_{\ell}\sin\delta_{p}\omega_{\ell}]\tran, \;(27) \nonumber
\end{align}

\vspace{-.2cm}\noindent
for $\ell=\{1,...,9\}$, $p=\{1,...,9\}$, and $p\ne\ell$; and Generator 10 is taken as a reference. The resulting multimachine power system of 18 state variables is lifted by 468 observable functions using EDMD-Lie, 190 observable functions using EDMD-p2, and 311 observable functions using EDMD-rbf311. In this case, EDMD-p3 (EDMD-p4) leads to 1,330 (7,315) observable functions and renders the KKF unobservable. This is because in the design of the filter, we assume that the following measurements are collected at the terminal of each generator: voltage phasor, current phasor, and real and reactive power. This assumption is consistent with the measurements provided by phasor measurement units. The available measurement set imposes a restriction on the number of observable functions that can be used. Moreover, the choice of observable functions also affects observability. Apart from the difference in the measurement set, the numerical simulation of the New England test system follows exactly the same strategy applied to the single-machine infinite-bus system. That is, 45 trajectories sampled in a linear region of the state space are recorded for 0.8 second, thereby leading to 160 samples per trajectory. These trajectories are used to compute the EDMD. Then, tests are performed on another set of trajectories sampled in a strongly nonlinear region of the state space. Note that the latter is not used to compute the EDMD. The accuracy in the prediction of a strongly nonlinear trajectory for each of the 18 state variables is shown in Fig. \ref{fig.multimachine}. In all tested cases, we observe that EDMD-Lie consistently outperforms EDMD-p2 and EDMD-rbf311.

\vspace{-.1cm}
\section{Conclusions and Future Work}
We provide an analytical method to select the observable functions to perform extended dynamic mode decomposition of nonlinear dynamics. This method is particularly attractive for dynamical systems where elementary nonlinear functions are beyond polynomial nonlinearities. The proposed method can be applied to a broad class of nonlinear dynamical systems encountered in many engineering fields. We demonstrate numerically that the proposed analytical method outperforms existing alternatives---that is, the use of monomials of order 2, 3, and 4, as well as the use of radial basis functions---in predicting strongly nonlinear trajectories. 

The application of the proposed method to control systems is ongoing research and of particular interest to power system models of increased complexity and size, including wind turbine models and their data-driven control. Given that there are Lie bracket-based guarantees for convergence of nonlinear controllers \cite{Mamakoukas2018}, this paper---by connecting the Koopman operator to Lie algebra---might be able to generate mathematical guarantees for the identification and control performance of controllable nonlinear systems using the proposed observable functions. A theoretical investigation of the interplay between the Koopman eigenfunctions and the basis obtained by polynomialization is also an interesting direction for future research.

\vspace{-.1cm}
\begin{appendices}
\section{Single-Machine Infinite-Bus System}\label{sec.appA}

\vspace{-.1cm}
Consider a synchronous generator represented by model 0.0, also referred to as the classical generator model:

\vspace{-.5cm}\begin{align}
\dot{\delta} &= \omega-\omega_{s}, \label{eq1a.SMIB.original} \\ 
M\dot{\omega} + \left(D/\omega_{s}\right)\left(\omega-\omega_{s}\right) &= P_{m} - P_{e}, \label{eq2a.SMIB.original}
\end{align}

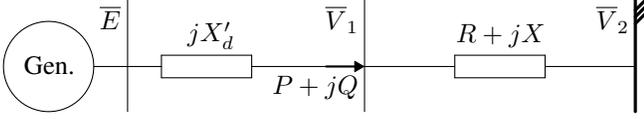
\begin{figure}[!h]
\centering
\begin{tikzpicture}[line cap=round,line join=round,>=triangle 45,x=1cm,y=1cm,scale=1.5]
\clip(8.3,3.55) rectangle (14.2,4.7);
\draw [line width=0.4pt] (8.8,4) circle (0.4cm);
\draw [line width=0.4pt] (9.5,4.6)-- (9.5,3.6);
\draw [line width=0.4pt] (11.6,4.6)-- (11.6,3.6);
\draw [line width=1.2pt] (14,4.6)-- (14,3.6);
\draw [line width=0.4pt] (13.2,4)-- (14,4);
\draw [line width=0.4pt] (12.4,4.1)-- (12.4,3.9);
\draw [line width=0.4pt] (12.4,3.9)-- (13.2,3.9);
\draw [line width=0.4pt] (13.2,3.9)-- (13.2,4.1);
\draw [line width=0.4pt] (13.2,4.1)-- (12.4,4.1);
\draw (9.17,4.62) node[anchor=north west] {$\overline{E}$};
\draw (11.17,4.62) node[anchor=north west] {$\overline{V}_{1}$};
\draw (13.57,4.62) node[anchor=north west] {$\overline{V}_{2}$};
\draw (9.95,4.48) node[anchor=north west] {$jX^{\prime}_{d}$};
\draw (12.33,4.46) node[anchor=north west] {$R+jX$};
\draw (10.7,4.00) node[anchor=north west] {$P+jQ$};
\draw (8.5,4.176831819712612) node[anchor=north west] {Gen.};
\draw [line width=1.2pt] (14.1,4.48)-- (14,4.38);
\draw [line width=1.2pt] (14.1,4.6)-- (14,4.5);
\draw [line width=1.2pt] (14.1,4.54)-- (14,4.44);
\draw [line width=0.4pt] (9.8,4.1)-- (9.8,3.9);
\draw [line width=0.4pt] (9.8,3.9)-- (10.6,3.9);
\draw [line width=0.4pt] (10.6,3.9)-- (10.6,4.1);
\draw [line width=0.4pt] (10.6,4.1)-- (9.8,4.1);
\draw [line width=0.4pt] (9.2,4)-- (9.8,4);
\draw [line width=0.4pt] (10.6,4)-- (12.4,4);
\draw [line width=0.8pt] (11.26,4)-- (11.54,4);
\begin{scriptsize}
\draw [fill=black,shift={(11.54,4)},rotate=270] (0,0) ++(0 pt,1.5pt) -- ++(1.299038105676658pt,-2.25pt)--++(-2.598076211353316pt,0 pt) -- ++(1.299038105676658pt,2.25pt);
\end{scriptsize}
\end{tikzpicture}
\vspace{-.7cm}
\caption{One-line diagram of the single-machine infinite-bus system adapted from Example 2.3 in \cite{Anderson2002}. $\overline{V}_{1}$, $\overline{V}_{2}$ are complex-valued voltage phasors in nodes $1$ and $2$, respectively. $P$ ($Q$) is the real (reactive) power injected into node $1$. $R$, $X$, and $X_{d}^{\prime}$ are parameters. We choose $R=0.05$, $X=0.30$, $V_{1}=1.05$, $V_{2}=1.00$, $P=0.80$, and $X^{\prime}_{d}=0.20$, all values in per unit.}
\label{fig.smib}
\end{figure}

\vspace{-.1cm}\noindent
connected to an infinite bus as in Fig. \ref{fig.smib}. In (\ref{eq1a.SMIB.original})--(\ref{eq2a.SMIB.original}), $\delta$ is the electrical angle related to the rotor mechanical angle, $\omega$ is the angular velocity of the revolving magnetic field, $\omega_{s}=120\pi$ is the synchronous angular velocity of the revolving magnetic field, $M=20/\omega_{s}$ is the inertia constant, $D=10$ is the damping constant, $P_{m}$ is the mechanical power, and $P_{e}$ is the electrical power.

\vspace{-.6cm}

\vspace{-.3cm}
Let $\overline{Y}=Ye^{j\gamma}=G+jB=\frac{1}{R+jX}$. Then:

\vspace{-.3cm}\begin{align}
\overline{S} =&\, P+jQ = \overline{V}_{1}\overline{I}^{*} = V_{1}^{2}Y\left(\cos\gamma-j\sin\gamma\right) \nonumber \\
& - V_{1}V_{2}Y\left[\cos\left(\theta_{1}-\gamma\right)+j\sin\left(\theta_{1}-\gamma\right)\right], \label{eq.S}
\end{align}

\vspace{-.3cm}\noindent
where $\overline{I}^{*}$ denotes the complex-conjugate of the current phasor injected into node $1$. Then:

\vspace{-.5cm}\begin{align}
P &= \text{Re}\left\{\overline{S}\right\} = V_{1}^{2}Y\cos\gamma - V_{1}V_{2}Y\cos\left(\theta_{1}-\gamma\right) \nonumber \\
&= V_{1}^{2}G - V_{1}V_{2}G\cos\theta_{1} - V_{1}V_{2}B\sin\theta_{1}. \label{eq.P}
\end{align}

\vspace{-.1cm}
By substituting the values of $V_{1}$, $V_{2}$, $G$, and $B$ into (\ref{eq.P}), and solving for $P=0.8$, one obtains $\theta_{1}=0.2243$. Then:

\vspace{-.5cm}\begin{align}
\overline{I} &= \left(\overline{V}_{1}-\overline{V}_{2}\right)\overline{Y} = 0.7718e^{j0.0640}. \nonumber \\
\overline{E} &= Ee^{j\delta} = \overline{V}_{1} + jX_{d}^{\prime}\overline{I} = 1.0854e^{j0.3651}. \nonumber
\end{align}

\vspace{-.1cm}
Let $\overline{S}_{e}=P_{e}+jQ_{e}$ and $\overline{Y}_{eq}=G_{eq}+jB_{eq}=\frac{1}{R+j\left(X+X_{d}^{\prime}\right)}$.

\noindent
Following (\ref{eq.S})--(\ref{eq.P}), we obtain $P_{e} = E^{2}G_{eq} - EV_{2}G_{eq}\cos\delta - EV_{2}B_{eq}\sin\delta$. Then:

\vspace{-.6cm}\begin{align}
\dot{\delta} &= \omega-\omega_{s}, \label{eq1a.SMIB} \\ 
M\dot{\omega} + a_{1}\left(\omega-\omega_{s}\right) &= a_{2} + a_{3}\cos\delta + a_{4}\sin\delta, \label{eq2a.SMIB}
\end{align}

\vspace{-.1cm}\noindent
where $a_{1} = \left(D/\omega_{s}\right)$, $a_{2} = P_{m} - E^{2}G_{eq}$, $a_{3} = EV_{2}G_{eq}$, $a_{4} = EV_{2}B_{eq}$, and the fixed point $\left(\delta_{0},\omega_{0}\right)=\left(0.3651,2\pi f\right)$. Now, to shift the fixed point to the origin, let $\delta=x_{1}+\delta_{0}$ and $\omega=x_{2}+\omega_{0}$. Then:

\vspace{-.5cm}\begin{align}
\dot{x}_{1} =&\, x_{2}+\omega_{0}-\omega_{s} = x_{2}. \nonumber \\
M\dot{x}_{2} +&\, a_{1}\left(x_{2}+\omega_{0}-\omega_{s}\right) = a_{2} + a_{3}\cos x_{1}\cos\delta_{0} \nonumber \\
-&\, a_{3}\sin x_{1}\sin\delta_{0} + a_{4}\sin x_{1}\cos\delta_{0} + a_{4}\sin\delta_{0}\cos x_{1}. \nonumber
\end{align}

\vspace{-.1cm}\noindent
Thus, $\dot{x}_{1} = x_{2}$, $\dot{x}_{2} = c_{00} + c_{0}x_{2} + c_{1}\cos x_{1} + c_{2}\sin x_{1}$, and $c_{00}=a_{2}/M$, $c_{0}=-a_{1}/M$, $c_{1}=\left(a_{3}\cos\delta_{0}+a_{4}\sin\delta_{0}\right)/M$, $c_{2}=\left(a_{4}\cos\delta_{0}-a_{3}\sin\delta_{0}\right)/M$. Note that the fixed point is now $\left(x_{1_{0}},x_{2_{0}}\right)=\left(0,0\right)$. By substituting values, $c_{00}=21.3649$, $c_{0}=-1$, $c_{1}=-21.3649$, and $c_{2}=-78.5772$.
\end{appendices}


\bibliographystyle{IEEEtran}

\end{document}